\documentclass[a4paper,11pt]{article}
\usepackage{res/jheppub}
\usepackage[T1]{fontenc}
\usepackage{color}

\title{\boldmath Cosmic Decoherence: Massive Fields}

\author[a,b]{Junyu Liu}
\author[c]{Chon-Man Sou}
\author[c]{Yi Wang}

\affiliation[a]{Department of Physics, California Institute of Technology, Pasadena, California 91125, USA}
\affiliation[b]{School of the Gifted Young, University of Science and Technology of China, Hefei, Anhui 230026, China}
\affiliation[c]{Department of Physics, The Hong Kong University of Science and Technology, Clear Water Bay, Kowloon, Hong Kong, China}

\emailAdd{jliu2@caltech.edu}
\emailAdd{cmsou@connect.ust.hk}
\emailAdd{phyw@ust.hk}

\abstract{We study the decoherence of massive fields during inflation based on the Zurek's density matrix approach. With the cubic interaction between inflaton and massive fields, the reduced density matrix for the massive fields can be calculated in the Schr\"odinger picture which is related to the variance of the non-Gaussian exponent in the wave functional. The decoherence rate is computed in the one-loop form from functional integration. For heavy fields with $m\gtrsim \mathcal{O}(H)$, quantum fluctuations will easily stay in the quantum state and decoherence is unlikely. While for light fields with mass smaller than $\mathcal{O}(H)$, quantum fluctuations are easily decohered within $5\sim10$ e-folds after Hubble crossing. Thus heavy fields can play a key role in studying problems involving inflationary quantum information.}

\begin{document}
\maketitle
\flushbottom

\section{Introduction}

The introduction of inflation resolves the horizon and flatness problems in standard big bang cosmology, and its direct consequence is to produce the large-scale cosmological perturbations \cite{Guth:1980zm,Linde:1981mu,Staro1982,Albrecht:1982wi,Hawking:1982cz,Guth:1982ec,GuthandPi1985}, leading to the anisotropies of cosmic microwave background (CMB) \cite{Planck2015a,Planck2015b,WMAP2013}. These perturbations can be traced back to the modes of quantum fluctuations, which left the horizon and became frozen during inflation and then reentered in radiation-or matter-dominated eras, and they seed the structure formations of the universe.
\\
\\
During inflation, the existence of massive fields (dubbed quasi-single field inflation \cite{Chen:2009we, ChenandWang2010,Chen2010,Baumann:2011nk, Assassi2012,Chen:2012ge,Noumi:2012vr,Gong:2013sma,Wang2014}) has many interesting properties theoretically and observationally. First, the presence of massive fields during inflation produces a ``quasi-local'' shape of non-Gaussianities. Moreover, scale-dependent massive fields can also generate particles during inflation \cite{Flauger:2016idt}, producing the cosmological collider physics signal through multi-correlation functions of primordial perturbations \cite{ArkaniandMaldacena2015, Chen:2016nrs, Lee:2016vti}, distinguishing different cosmological models by the quantum primordial standard clock effects \cite{ChenNamjooWang2016,Probing2016,Chen:2016qce}.
\\
\\
Recently, there is a growing interest in studying the quantum information contained in the cosmological perturbations \cite{Lim:2014uea,Maldacena:2015bha,Choudhury:2016cso,
Markkanen:2016aes,Kanno:2016gas}. For this purpose, it is important to make sure that decoherence has not happened, so that the quantum fluctuations have not become classical. Then we need to answer a question, is massive field in the early universe in the quantum state, or it is decohered by the interaction with curvature perturbation? In this paper, we will address in detail the decoherence problem of massive field during inflation.
\\
\\
For curvature perturbation of inflaton, it is believed that wave modes of quantum fluctuations will be frozen as classical perturbation soon after the Hubble exit and be successfully decohered. Previous research on the decoherence of cosmological perturbations investigated several aspects: (i) The squeezing behavior of the quantum state at late time caused by a large particle occupation number and a large squeezed parameter (that can be called \emph{squeezed state}) \cite{Albrecht1994,Lesgourgues:1996jc,Kiefer:1998qe,GuthandPi1985}, makes the non-commutativity between field operator and its conjugate momentum vulnerable, and therefore the statistical behavior is closed to a classical distribution. This argument suggests that classicality can emerge easily in inflationary scenario, but actual decoherence does not really appear in this framework because the essential interactions have not been considered yet, and thus it is called \emph{decoherence without decoherence}. (ii) The Markovian approximation \cite{Kiefer:2006je,Kiefer:2008ku,Burgess:2006jn,Burgess:2014eoa,Burgess:2015ajz}, in which the evolution of density matrix is governed by the master equation with Lindblad operator \cite{book2003}, estimates the decoherence caused by interactions between system and environment. One noteworthy thing is that stochastic inflation with noise and drift may appear due to interactions \cite{Burgess:2014eoa,Burgess:2015ajz}, and particularly the interaction with massive field may produce colored noise which affects CMB anisotropy \cite{Wu:2006xp}. This kind of methods assumes the correlation time between environment and system is much shorter than the interaction time-scale, but practically one has to consider specific type of interactions in cosmological problems instead of a generic study. (iii) Decoherence with actual interactions, including interaction with isocurvature perturbations \cite{ProkopecandRigopoulos2007}, cubic interaction with vacuum \cite{Koksma2010} and thermal bath \cite{Koksma2011} respectively, and gravitational nonlinearities \cite{Nelson:2016kjm}. The last method corresponds to the original understanding of quantum decoherence by celebrated series of works of Zurek \cite{Zurek01,Zurek02,Zurek1,Zurek2,Zurek3,Zurek4} without further approximations. In this approach, one splits system and environment, considers their entanglement and interaction during the quantum evolution, and traces out the environment to get a reduced density matrix with decreasing contribution of off-diagonal terms which labels quantum decoherence. In this work, we will mainly use this method to investigate the massive field decoherence by choosing nontrivial interaction between inflaton and a massive field and choosing inflaton as the environment while the massive field as the system.
\\
\\
In this paper, we will investigate in detail on the decoherence of massive fields during inflation. With the decoherence formalism by Zurek, we obtain that the decoherence rate changes with conformal time, the wave number, the coupling constant of interaction and the mass of massive field. Assuming proper coupling range related to current observations \cite{Planck2015c,AssassiandBaumann2014}, we find that for mass smaller than $\mathcal{O}(H)$, quantum fields are easily decohered within $5\sim10$ e-folds after crossing horizon. However, for mass larger than $\mathcal{O}(H)$, and especially for very massive field $m\gg \mathcal{O}(H)$, quantum fluctuations are very hard to be decohered and they are expected to stay in the quantum state during inflation.
\\
\\
This paper is organized as follows. In Section \ref{without}, we extend the discussion on ``decoherence without decoherence'' to massive fields. Namely we compute the one-mode occupation number for massive fields. In Section \ref{deco}, we focus on the actual decoherence of massive fields by the reduced density matrix approach. Firstly, we will review the theoretical construction of quantum decoherence following Zurek. Secondly, a generic formalism of quantum evolution considering system and environment is introduced based on the previous study \cite{Nelson:2016kjm}. Thirdly, the reduced density matrix is computed and through functional integration, we give the one-loop representation of decoherence rate. Then, we evaluate the loop integration both analytically and numerically, and the likelihood of cosmic decoherence with different masses are shown quantitatively. We conclude in Section \ref{conclu}. In this paper, we set the reduced Planck mass $M_\text{pl}=1$.

\section{Curved spacetime: squeezed state or not?}\label{without}

For inflationary cosmology with a massless curvature mode only, the well-know cosmic decoherence theory predicts a strongly squeezed state near the Hubble crossing. This squeezed state is produced by the strongly curved spacetime geometry, which causes the dynamical mixing of canonical fields and their conjugate momenta, and the modification of the creation annihilation operators by a time-dependent Bogolubov transformation. As a result, the one-mode occupation number $n_k$ diverges in de Sitter spacetime. This phenomena causes a largely enhanced possibility of actual quantum decoherence in the understanding of Zurek \cite{Zurek01,Zurek02,Zurek1,Zurek2,Zurek3,Zurek4}. Namely, a large occupation number means a system which is approximately classical, something like Schr\"odinger's cat instead of a fundamental particle theory system with low particle numbers. A large occupation number is related to a squeezed parameter \cite{Polarski:1995jg,Lesgourgues:1996jc,Kiefer:1998jk,Kiefer:1998qe}, $r_k$, as
\begin{align}
n_k = \sinh^2 (r_k)~,
\end{align}
where the delta function has been factored out. Large $n_k$ or large $r_k$ implies the semiclassicality of the system, with a large number of particles and a strongly suppressed commutation relation. This gives a squeezed phase space \cite{Kiefer:1998jk} (that is why it is called \emph{squeezed}).
\\
\\
This phenomena is known as \emph{decoherence without decoherence} and has been discussed in a series of studies (eg. \cite{Kiefer:2008ku,Polarski:1995jg,Lesgourgues:1996jc,Kiefer:1998qe,Mijic:1998if}). Decoherence without decoherence is strongly connected to the actual decoherence by some mathematical and physical arguments both in cosmology and quantum lab condition. This connection can be seen from the positivity of Wigner function with the full density matrix, where the time for a positive Wigner function (as a feature for a nearly classical state) is approximately the time for actual decoherence because of a large possibility to be quickly decohered as a squeezed state \cite{Kiefer:2006je,math1,math2}.
\\
\\
In this section, we will introduce the one-mode occupation number in the context for general mass. Unlike the massless case, a massive field in de Sitter behaves differently in the context of decoherence without decoherence. From this analysis, one can get the physical intuition to determine in which situation decoherence takes place (having in mind that interaction delays the actual decoherence time, as we shall see in later sections). Part of this computation is achieved by some previous works \cite{Polarski:1995jg,Mijic:1998if}.
\\
\\
Let us consider the Lagrangian for a massive field in de Sitter space,
\begin{align}
{\mathcal{L}_\sigma}=\sqrt{-g}(-\frac{1}{2}{{\partial }^{\mu }}\sigma {{\partial }_{\mu }}\sigma -\frac{1}{2}{{m}^{2}}{{\sigma }^{2}})~,
\end{align}
the time-dependent quantum solution for the mode function is given by
\begin{align}
\sigma (\tau ,\mathbf{x})=\int{\frac{{{d}^{3}}\mathbf{k}}{{{(2\pi )}^{3}}}{{e}^{i\mathbf{k}\cdot \mathbf{x}}}}({{\varsigma}_{k}}(\tau ){{a}_{\mathbf{k}}}(\tau_i)+\varsigma_{k}^{*}(\tau )a_{-\mathbf{k}}^{\dagger }(\tau_i))~,
\end{align}
where
\begin{align}
{{\varsigma}_k}(\tau )=i{{e}^{\frac{\pi i}{4}+\frac{\pi i\nu }{2}}}\frac{\sqrt{\pi }}{2}H{{(-\tau )}^{3/2}}H_{\nu }^{(1)}(-k\tau )~,
\end{align}
and $\nu =\sqrt{\frac{9}{4}-\frac{{{m}^{2}}}{{{H}^{2}}}}$ (note that the index $\nu$ can be a pure imaginary number if $m$ is larger than $\frac{3H}{2}$). We denote $\tau$ is the conformal time from $-\infty$ to $0$, and $\tau_0$ for a specific initial time of inflation. If we define a new field variable $y=a\sigma$, the new field
\begin{align}
 y(\tau ,\mathbf{x})=\int{\frac{{{d}^{3}}\mathbf{k}}{{{(2\pi )}^{3}}}{{e}^{i\mathbf{k}\cdot \mathbf{x}}}}({{z}_{k}}(\tau ){{a}_{\mathbf{k}}}(\tau_0)+z_{k}^{*}(\tau )a_{-\mathbf{k}}^{\dagger }(\tau_0))~,
\end{align}
with
\begin{align}
{{z}_{k}}(\tau )=i{{e}^{\frac{\pi i}{4}+\frac{\pi i\nu }{2}}}\frac{\sqrt{\pi }}{2}{{(-\tau )}^{1/2}}H_{\nu }^{(1)}(-k\tau )~,
\end{align}
corresponds to ordinary quantized scalar field in the flat space quantum field theory. In the subhorizon limit, the mode function has a consistent behavior with the flat space case, namely the Bunch-Davis vacuum
\begin{align}
{{z}_{k}}(\tau \to {{\tau }_{0}})=\frac{1}{\sqrt{2k}}{{e}^{-ik\tau }}~.
\end{align}
One can also write down the Lagrangian in terms of field $y$
\begin{align}
\mathcal{L}_\sigma=\frac{1}{2}{{a}}{{\dot{y}}^{2}}-\frac{1}{2a}{{\partial }_{i}}y{{\partial }_{i}}y+\frac{1}{2}(\frac{{{{\dot{a}}}^{2}}}{a}-{{m}^{2}}a){{y}^{2}}-\dot{a}\dot{y}y~,
\end{align}
and also, the Hamiltonian
\begin{align}
\mathcal{H}_\sigma=\frac{{\dot{a}}}{a}\Pi y+\frac{1}{2}\frac{{{\Pi }^{2}}}{a}+\frac{1}{2a}{{\partial }_{i}}y{{\partial }_{i}}y+\frac{a}{2}{{m}^{2}}{{y}^{2}}~,
\end{align}
with the Lengendre transformation
\begin{align}\label{momen}
\Pi =y'-\frac{a'}{a}y~,
\end{align}
where primes mean the derivative for the conformal time coordinate. Thus, for the following one-mode field and its conjugate
\begin{align}
&{{y}_{\mathbf{k}}}(\tau )={{z}_{k}}(\tau ){{a}_{\mathbf{k}}}({{\tau }_{0}})+z_{k}^{*}(\tau )a_{-\mathbf{k}}^{\dagger }({{\tau }_{0}})~,\nonumber\\
&{{\Pi }_{\mathbf{k}}}(\tau )=\left( {{z}_{k}'}(\tau )-\frac{a'}{a} {{z}_{k}}(\tau )\right){{a}_{\mathbf{k}}}({{\tau }_{0}})+\left( z_{k}'^{*}(\tau )-\frac{a'}{a}{{z}^*_{k}}(\tau ) \right)a_{-\mathbf{k}}^{\dagger }({{\tau }_{0}})~,
\end{align}
the time-dependent creation annihilation operators in the Heisenberg picture will be given by
\begin{align}
  & {{a}_{\mathbf{k}}}(\tau )=\frac{1}{\sqrt{2}}\left( \sqrt{\left| {{\omega }_{k}}(\tau ) \right|}{{y}_{\mathbf{k}}}(\tau )+\frac{i}{\sqrt{\left| {{\omega }_{k}}(\tau ) \right|}}{{\Pi }_{\mathbf{k}}}(\tau ) \right)~,  \nonumber\\
 & a_{-\mathbf{k}}^{\dagger }(\tau )=\frac{1}{\sqrt{2}}\left( \sqrt{\left| {{\omega }_{k}}(\tau ) \right|}{{y}_{\mathbf{k}}}(\tau )-\frac{i}{\sqrt{\left| {{\omega }_{k}}(\tau ) \right|}}{{\Pi }_{\mathbf{k}}}(\tau ) \right) ~,
\end{align}
where the frequency is given as
\begin{align}
{{\omega }_{k}}(\tau )=\sqrt{{{m}^{2}}{{a}^{2}}+{{k}^{2}}-\frac{{{a}''}}{a}}=k\sqrt{\frac{1}{{{k}^{2}}{{\tau }^{2}}}\left( \frac{1}{4}-{{\nu }^{2}} \right)+1}~.
\end{align}
This is essentially the Bogolubov transformation. Thus the one-mode occupation number (factored out the delta function) is given by
\begin{align}\label{nk}
{{n}_{k}}(\tau )&=\frac{1}{2}{{\left| \sqrt{\left| {{\omega }_{k}}(\tau ) \right|}{{z}_{k}}(\tau )-\frac{i}{\sqrt{\left| {{\omega }_{k}}(\tau ) \right|}}\left( {{{{z}}}'_{k}}(\tau )-\frac{{{a}'}}{a} z_{k}(\tau )\right) \right|}^{2}}\nonumber\\
&=\frac{1}{2}\left( \left| {{\omega }_{k}}(\tau ) \right|{{z}_{k}}(\tau )z_{k}^{*}(\tau )+\frac{1}{\left| {{\omega }_{k}}(\tau ) \right|}{{\left| {{z}_{k}'}(\tau )-\frac{{{a}'}}{a} \right|}^{2}}+2\operatorname{Im}z_{k}^{*}(\tau )\left( {{z}_{k}'}(\tau )-\frac{{{a}'}}{a} \right) \right)~.
\end{align}
It's difficult to see the behavior of occupation number as functions of $-k\tau$ and $m$ from Hankel functions directly.
Thus we are going to use the Taylor expansion to show its behavior around $k\tau\to0^-$ with the expansion formula of Hankel functions. Different leading order scalings with real, imaginary cases of $\nu$, and different regions below or above the conformal mass $\nu^2=\frac{1}{4}$ will cause totally different situations of $n_k(\tau)$, namely
\begin{itemize}
\item For $0<\nu<\frac{1}{2}$, or $\sqrt{2}H<m<\frac{3}{2}H$, we obtain
\begin{align}
{{n}_{k}}(\tau )= \frac{{{2}^{-3+2\nu}}(5-6\nu)\Gamma^2 {{(\nu)}}}{\pi \sqrt{1-4{{\nu}^{2}}}}{{(-k\tau )}^{-2\nu}}~,
\end{align}
thus a larger mass will cause a smaller occupation number in this region, namely harder to decohere.
\item For $\nu=\frac{1}{2}$, namely $m=\sqrt{2}H$ for the conformal mass, we obtain
\begin{align}
{{n}_{k}}(\tau )= \frac{1}{4}{{(-k\tau )}^{-2}}~.
\end{align}
The reason is given as follows. Although in the conformal mass the mode function is the same as the flat spacetime. However, the conjugate field momentum is still modified by de Sitter space. The gravitational contribution in the momentum (Equation (\ref{momen})) gives the modification term $\frac{a'}{a}z_k$ in Equation (\ref{nk}). Although the first two terms in the first line of Equation (\ref{nk}) cancel each other, the $\frac{a'}{a}z_k$ term gives a contribution $\frac{1}{2}{{\left| \frac{i}{\sqrt{k}}\frac{{{a}'}}{a}{{z}_{k}}(\tau ) \right|}^{2}}=\frac{1}{4}{{(-k\tau )}^{-2}}$. Thus, in de Sitter the scalar field with conformal mass is still in the squeezed state near exiting the horizon.
\item For $\frac{1}{2}<\nu \le \frac{3}{2}$, or $0 \le m < \sqrt{2}H$, we obtain
\begin{align}
{{n}_{k}}(\tau )= \frac{{{2}^{-2+2\nu}}(2-3\nu+2{{\nu}^{2}})\Gamma^2 {{(\nu)}}}{\pi \sqrt{4{{\nu}^{2}}-1}}{{(-k\tau )}^{-2\nu}}~.
\end{align}
Thus we know that the occupation number for conformal and non-conformal case is not continuous because the dependence comes to another branch in different regions of $\nu$.
\item For $\nu =0 $, or $m=\frac{3}{2}H$, we obtain
\begin{align}
{{n}_{k}}(\tau )= \frac{5{{\log }^{2}}(-k\tau )}{2\pi }~.
\end{align}
In this critical damping case we get a logarithmic divergence which is still a squeezed state.
\item For $\nu \in i\mathbb{R}^+ $, or $m>\frac{3}{2}H$, defining $\nu=iv$ with $v>0$, we have
\begin{align}
{{n}_{k}}(\tau )=\frac{(5+4{{v}^{2}})\coth (\pi v)}{4v\sqrt{1+4{{v}^{2}}}}-\frac{1}{2}+\frac{1}{4\pi \sqrt{1+4{{v}^{2}}}}\operatorname{Re}\left[ {{2}^{-2iv}}(5+6iv){{\Gamma }^{2}}(-iv){{(-k\tau )}^{2iv}} \right]~.
\end{align}
This result shows a vibrating feature for very massive field in de Sitter space. There is no divergence near dS boundary $k\tau\to0^+$. In fact, the result is vibrating but bounded by
\begin{align}
{{n}_{k}}(\tau )\le \vartheta (v)=\frac{(5+4{{v}^{2}})\coth (\pi v)}{4v\sqrt{1+4{{v}^{2}}}}-\frac{1}{2}+\frac{\left| \Gamma ^2 (iv) \right|\sqrt{25+36v}}{4\pi \sqrt{1+4{{v}^{2}}}}~.
\end{align}
One can plot the bounded function as the following Fig.~\ref{without1}. One can see that the bounded function approaches zero. Quantitatively, we have the large mass expansion
\begin{figure}[htbp]
  \centering
  \includegraphics[width=0.9\textwidth]{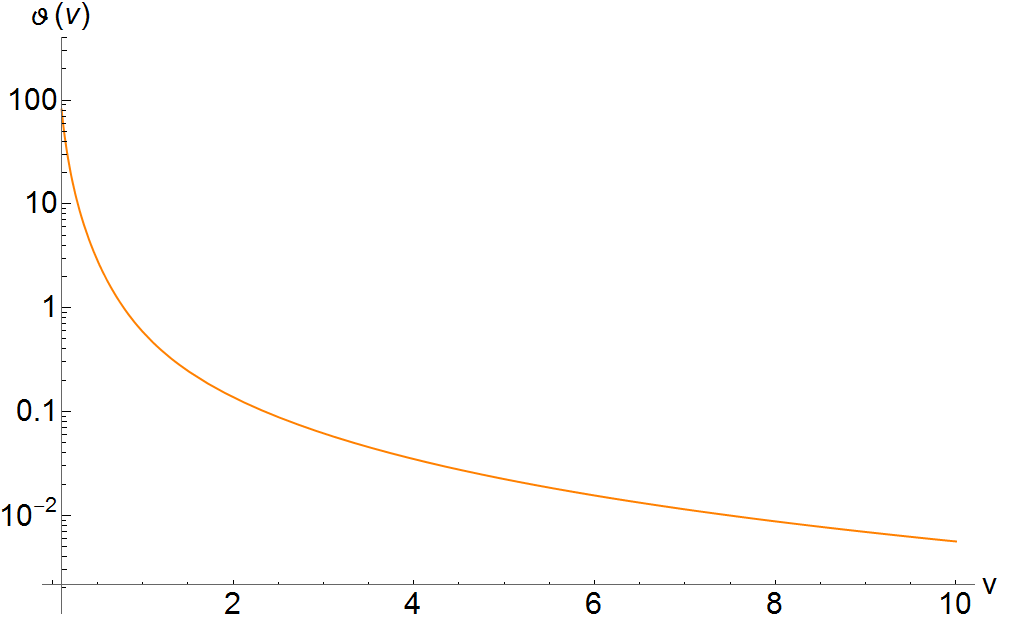}
  \caption{\label{without1} The function $\vartheta(v)$ as the maximal ${{n}_{k}}(\tau )$. We use the logarithmic scaling for the function $\vartheta (v)$.}
\end{figure}
\begin{align}
\vartheta (v)\sim \frac{9}{16{{v}^{2}}}\coth [\pi v]\sim \frac{9}{16{{v}^{2}}}~.
\end{align}
So the vibration amplitude of occupation number will decrease with the increasing mass.
\end{itemize}
To summarize for the one-mode occupation number for different mass in de Sitter space, now we can successfully distinguish whether a given scalar field will occupy a squeezed state in de Sitter space, which is given as follows,
\\
\begin{itemize}
\item Squeezed state ($0\le m \le \frac{3}{2} H$):\\
In this mass region the quantum fields will occupy a divergent occupation number near the Hubble crossing, in which a process called decoherence without decoherence will happen and gives a large possibility to achieve the actual decohere. Generally, the divergent scale will be $(-k\tau)^{-2\nu}$ where $\nu =\sqrt{\frac{9}{4}-\frac{{{m}^{2}}}{{{H}^{2}}}}$. The only two exceptions will be the conformal mass $m=\sqrt{2}H$ or $\nu=\frac{1}{2}$ (which suddenly skips different branches of the dependence and gives a divergence scaling $(-k\tau)^{-2}$) and the critical damping mass $m=\frac{3}{2}H$ or $\nu=0$ (which will gives a logarithmic divergence $\log(-k\tau)$).
\end{itemize}
\begin{itemize}
\item Non-squeezed state or quantum state ($m > \frac{3}{2} H$):\\
In this mass region the quantum fields will never occupy the infinity particle number during inflation. Near the Hubble crossing epoch quantum fields will vibrate very fast as $\sin(\log(-k\tau))$. However, this vibration will be suppressed by $(\text{Im}[\nu])^{-2}$, thus the particle number will be smaller for a larger mass, and harder to decohere. Also, for a super massive field $m\approx M_\text{pl}=1$ and even larger, the particle number will easily fall to zero.
\end{itemize}

\section{Actual quantum decoherence for general mass}\label{deco}

\subsection{Towards a classical universe}\label{decoherence theory}

After introducing decoherence without decoherence, we give a short introduction on the quantum decoherence following Zurek \cite{Zurek01,Zurek02,Zurek1,Zurek2,Zurek3,Zurek4}.
\\
\\
In Schr\"odinger picture, a system is described by an evolving quantum state that can be expressed by various superpositions with different bases. For clearly showing the probability interpretation of a quantum state, a pure state can define a density matrix with basis
\begin{align}
|\psi\rangle=\sum_{i}\Psi_{i}|b_{i}\rangle\Rightarrow\rho=|\psi\rangle\langle\psi|=\sum_{i}\Big|\Psi_{i}\Big|^{2}|b_{i}\rangle\langle b_{i}|+\sum_{i\neq j}\Psi_{i}\Psi_{j}^{*}|b_{i}\rangle\langle b_{j}|\label{eq:density matrix}~.
\end{align}
The diagonal terms are interpreted as probability density, and the off-diagonal terms are ascribed to the quantum coherence of state. Such a quantity includes the statistical properties needed for describing the system. On the other hand, the interaction with environment, which is defined by the part outside the system, can select untouched, preferred pointer states to form a particular basis after time evolution, according to the theory of environment-induced superselection (Einselection) \cite{Zurek01,Zurek02,Zurek3,Zurek4},
\begin{align}
|\psi\rangle|\mathcal{E}\rangle\rightarrow\sum_{i}\tilde{\Psi}_{i}|a_{i}\rangle|\mathcal{E}_{i}\rangle\ ~~~,~~~ \langle\mathcal{E}_{i}|\mathcal{E}_{j}\rangle\approx\delta_{ij}\label{eq:enviroment}~.
\end{align}
The nearly orthogonality are attributed to the numerous degrees of freedom of environment, and this condition is essential for quantum decoherence process. The composite of system and environment forms a Hilbert space via tensor product of two sub-Hilbert spaces,
\begin{align}
\mathcal{H_T}=\mathcal{H_{S}}\otimes\mathcal{H_{E}}~.
\end{align}
Physically, all the observations of system are only related to the knowledge acquired from the system instead of the whole composite. Mathematically, the expectation values of such observations depend on the density matrix after tracing out the environment,
\begin{align}
\langle\mathcal{O_{S}}\otimes I_{\mathcal{E}}\rangle=\text{Tr}_{\mathcal{S}}\Big(\text{Tr}_{\mathcal{E}}(\rho)\mathcal{O_{S}}\Big)~.
\end{align}
The operator $\rho_{R}=\text{Tr}_{\mathcal{E}}(\rho)$ is called the reduced density matrix, and clearly it records the statistical properties including probability and quantum coherence, as shown in Equation (\ref{eq:density matrix}). With the process of Einselection and orthogonality of environment states shown in Equation (\ref{eq:enviroment}), the reduced density matrix is diagonalized,
\begin{align}
\rho_{R}=\sum_{i}\Big|\tilde{\Psi}_{i}\Big|^{2}|a_{i}\rangle\langle a_{i}|\label{eq:diagonal}~.
\end{align}
The matrix with such a form defines a mixed state in which possible configurations have classical probabilities $p_{i}=\Big|\tilde{\Psi}_{i}\Big|^{2}$, and the expectation values of observations are expressed by ensemble average,
\begin{align}
\langle\mathcal{O_{S}}\rangle=\sum_{i}p_{i}\langle a_{i}|\mathcal{O_{S}}|a_{i}\rangle~.
\end{align}
Comparing with Equation (\ref{eq:density matrix}), the terms with quantum coherence disappear, and the reduced density matrix only records classical statistics. Such phenomenon can also be understood as loss of the information that is quantified by the increased entropy of the mixed state \cite{Zurek4},
\begin{align}
S_{\text{mixed}}=-\text{Tr}_{\mathcal{S}}\Big(\rho_{R}\log\rho_{R}\Big)>S_{\text{pure}}=0~,
\end{align}
leading to quantum decoherence of system states. Overall, the decoherence is characterized by the vanish off-diagonal terms in the reduced density matrix with the basis formed by pointer states. Regarding the pointer states of quantum fluctuations during inflation, previous research had argued that the environment distinguishes the field-amplitude basis \cite{Kiefer:1998qe,Kiefer:2006je}, and thus it is proper to analyze the wave functional of the whole composite of the massive field and the inflaton. With the field-amplitude representation, the trace of density matrix is given by functional integration with respect to inflaton, and thus the reduced density matrix can be calculated if the wave functional is obtained.

\subsection{Settings and a generic formalism}

We consider the theory including a (nearly) massless inflaton $\varphi$ and a massive field $\sigma$ during the course of inflation. The interaction which renders the inflaton field to decohere the massive field is a dimension-five operator.
\begin{align}\label{eq:Largrian interaciton}
\mathcal{L}_{\text{int}}=-ga\sigma(\partial_{i}\varphi)(\partial_{i}\varphi)=-\frac{a}{2}\frac{\sigma(\partial_{i}\varphi)(\partial_{i}\varphi)}{\Lambda}
~,
\end{align}
This kind of interaction appears in the effective field theory (EFT) of multi-field inflation, and the observations constrain the cutoff scale to satisfy $\Lambda>\mathcal{O}(10^{-3}\sim1)M_{\text{pl}}=\mathcal{O}(10^{-3}\sim1)$ \cite{AssassiandBaumann2014}. In \cite{Nelson:2016kjm}, the coupling constant (for inflaton) is
\begin{align}
g_{\text{GR}}=\frac{(\epsilon+\eta)}{4\sqrt{2\epsilon}M_{\text{pl}}}<\frac{1}{\Lambda}~,
\end{align}
which is comparable to the case of EFT, hence we choose this to be the coupling constant. Therefore, the interaction is assumed to be small. With this interaction, we can study the evolution of density matrix in the inflationary background. It is useful to calculate the interaction Hamiltonian,
\begin{align}\label{eq: interaction Hamiltonian in momentum space}
H_\text{int}&=g\int d^{3}\mathbf{x}\ a(\tau)\sigma(\partial_{i}\varphi)(\partial_{i}\varphi)\nonumber\\
&=-g\int d^{3}\mathbf{x}\int\frac{d^{3}\mathbf{k}}{(2\pi)^{3}}\frac{d^{3}\mathbf{k'}}{(2\pi)^{3}}\frac{d^{3}\mathbf{q}}{(2\pi)^{3}}\ e^{i(\mathbf{k}+\mathbf{k}'+\mathbf{q})\cdot\mathbf{x}}a(\tau)\varphi_{\mathbf{k}}\varphi_{\mathbf{k}'}\sigma_{\mathbf{q}}(\mathbf{k}\cdot\mathbf{k}')\nonumber\\
&=\int_{\mathbf{k,k',q}}\varphi_{\mathbf{k}}\varphi_{\mathbf{k}'}\sigma_{\mathbf{q}}\tilde{\mathcal{H}}_{\mathbf{k,k',q}}^{(\text{int})}(\tau)~,
\end{align}
where
\begin{align}
\tilde{\mathcal{H}}_{\mathbf{k,k',q}}^{(\text{int})}(\tau)=\frac{g}{2H\tau}(q^{2}-k^{2}-k'^{2})~,
\end{align}
is calculated with Fourier transform and governs the interaction in momentum space. The integration symbol means
\begin{align}
\int_{\mathbf{k},\mathbf{k}',\mathbf{q}}=\int\frac{d^{3}\mathbf{k}}{(2\pi)^{3}}\frac{d^{3}\mathbf{k'}}{(2\pi)^{3}}\frac{d^{3}\mathbf{q}}{(2\pi)^{3}}(2\pi)^{3}\delta^{3}(\mathbf{k}+\mathbf{k'}+\mathbf{q})~.
\end{align}
The quantum state, which involves both two quantum field, evolves with the sum of free  and interaction Hamiltonians,
\begin{align}
H&=H_{0}+H_{\text{int}}\nonumber\\
&= \frac{1}{2}\int\frac{d^{3}\mathbf{k}}{(2\pi)^{3}}\ \Big\{\frac{\pi_{\mathbf{k}}^{(\varphi)}\pi_{\mathbf{-k}}^{(\varphi)}}{a(\tau)^{3}}+a(\tau)k^{2}\varphi_{\mathbf{k}}\varphi_{-\mathbf{k}}+\frac{\pi_{\mathbf{k}}^{(\sigma)}\pi_{-\mathbf{k}}^{(\sigma)}}{a(\tau)^{3}}+\Big[a(\tau)k^{2}+a(\tau)^{3}m^{2}\Big]\sigma_{\mathbf{k}}\sigma_{-\mathbf{k}}\Big\}\nonumber\\
&+\int_{\mathbf{k,k',q}}\varphi_{\mathbf{k}}\varphi_{\mathbf{k}'}\sigma_{\mathbf{q}}\tilde{\mathcal{H}}_{\mathbf{k,k',q}}^{(\text{int})}(\tau)\label{eq:Hamiltonain}~.
\end{align}
The interaction part produces the non-Gaussianities to the total wave functional that leads to vanish off-diagonal terms of reduced density matrix. Without using the ansatz $\Psi[\varphi,\sigma]=\Psi_{\text{G}}^{(\varphi)}[\varphi]\Psi_{\text{G}}^{(\sigma)}[\sigma]\Psi_{\text{NG}}[\varphi,\sigma]$
in \cite{Nelson:2016kjm}, we calculate time-evolution operators to find the non-Gaussian part of wave functional. Firstly, we define operator
\begin{align}
G(\tau)=U(\tau,\tau_{0})U_{0}^{-1}(\tau,\tau_{0})~,
\end{align}
where $U$ and $U_0$ are ordinary unitary operators defined in QFT, and $\tau_0$ is  initial conformal time. Thus, the operator $G$ transforms the state in free field theory to interaction theory. With the Schr\"odinger equations of free and interaction theories, the differential equation for $G(\tau)$ is obtained as
\begin{align}
&\begin{cases}
i\frac{\partial|\psi_{\text{G}}\rangle}{\partial t}=H_{0}[\varphi(\tau_{0}),\sigma(\tau_{0}),\tau]|\psi_{\text{G}}\rangle\\
i\frac{\partial|\psi\rangle}{\partial t}=H[\varphi(\tau_{0}),\sigma(\tau_{0}),\tau]|\psi\rangle
\end{cases}\nonumber\\
&\Rightarrow\frac{\partial G}{\partial\tau}=-ia(\tau)\Big\{[H[\varphi(\tau_{0}),\sigma(\tau_{0}),\tau],G]+GH_{\text{int}}[\varphi(\tau_{0}),\sigma(\tau_{0}),\tau]\Big\}~,\label{eq:equation for G}
\end{align}
where $|\psi_{\text{G}}\rangle$ and $|\psi\rangle$ are the quantum states in free field and interaction theories respectively. The initial state at $\tau_{0}$ is assumed to be equal,
\begin{align}
|\psi(\tau_{0})\rangle=|\psi_{\text{G}}(\tau_{0})\rangle~,
\end{align}
and thus they satisfy
\begin{align}
|\psi(\tau)\rangle=G(\tau)|\psi_{\text{G}}(\tau)\rangle~.
\end{align}
It is noticeable that the field arguments inside the Hamiltonians are the fields at initial time that are different from those in Heisenberg and interaction pictures. With the result from interaction picture, the solution of $G(\tau)$ is easily found,
\begin{align}\label{eq:G from interaction picture}
G(\tau)&=U(\tau,\tau_{0})\Big(U_{0}^{-1}(\tau,\tau_{0})U(\tau,\tau_{0})\Big)U^{-1}(\tau,\tau_{0})\nonumber\\
&=\mathbb{T}\exp{\left(-i\int_{\tau_{0}}^{\tau}d\tau'\ a(\tau')U(\tau,\tau')H_{\text{int}}[\varphi(\tau_{0}),\sigma(\tau_{0}),\tau']U^{-1}(\tau,\tau')\right)}~,
\end{align}
where $\mathbb{T}$ is the time-ordered operator. It is easy to verify that such an expression satisfies Equation (\ref{eq:equation for G}). The general form of this time-ordered exponential is difficult to calculate. But with the assumption that interaction is weak enough, the result is simply related to leading terms. On the other hand, the Gaussian state to which $G(\tau)$ operates can be solved exactly, and its wave functional is given by the product of two independent Gaussian wave packets \cite{Burgess:2014eoa},
\begin{align}\label{eq:Gaussian wave packet}
\Psi_{\text{G}}[\varphi,\sigma]=\Psi_{\text{G}}^{(\varphi)}[\varphi]\Psi_{\text{G}}^{(\sigma)}[\sigma]=\mathcal{N}(\tau)\exp\Big[-\int\frac{d^{3}\mathbf{k}}{(2\pi)^{3}}\Big(\varphi_{\mathbf{k}}\varphi_{-\mathbf{k}}A_{\varphi}+\sigma_{\mathbf{k}}\sigma_{-\mathbf{k}}A_{\sigma}\Big)\Big]~,
\end{align}
where the factor inside the exponent can be calculated with the Schr\"odinger equation,
\begin{align}
A_{\varphi}(k,\tau)=-\frac{i}{(H\tau)^{2}}\frac{u'_{k}(\tau)}{u_{k}(\tau)} ~~~,~~~A_{\sigma}(k,\tau)=-\frac{i}{(H\tau)^{2}}\frac{w'_{k}(\tau)}{w_{k}(\tau)}~,\label{eq:A}
\end{align}
and the (reduced) mode functions are written to be consistent with the conventions of \cite{Burgess:2014eoa,Nelson:2016kjm},
\begin{align}
u_{k}(\tau)\propto(-k\tau)^{\frac{3}{2}}H_{\frac{3}{2}}^{(2)}(-k\tau)~~~,~~~w_{q}(\tau)\propto(-q\tau)^{\frac{3}{2}}H_{\nu}^{(2)}(-q\tau)~.
\end{align}
The mode function $u_k$ represents the quantum fluctuation solution for the massless inflaton with a Hankel index $\frac{3}{2}$, while the mode function $w_q$ represents the solution for massive field with a generic Hankel index $\nu=\sqrt{\frac{9}{4}-\frac{m^{2}}{H^{2}}}$. For inflaton, the Hankel function can be reduced explicitly into trianglar functions, and the late-time expansion of $A_{\varphi}(k,\tau)$ is given by
\begin{align}
A_{\varphi}(k,\tau)\approx-\frac{ik^{2}}{H^{2}\tau}+\frac{k^{3}}{H^{2}}+\mathcal{O}(k\tau)~,
\end{align}
which means that the width approaches to a constant and the phase changes rapidly. With this field-amplitude representation of the Gaussian state, the first order non-Gaussianities generated by $G(\tau)$ can be calculated. Firstly, we define an operator
\begin{align}
\mathcal{K}_{\mathbf{k,k',q}}(\tau',\tau)=U(\tau,\tau')\varphi_{\mathbf{k}}\varphi_{\mathbf{k'}}\sigma_{\mathbf{q}}U^{-1}(\tau,\tau')=U^{-1}(\tau',\tau)\varphi_{\mathbf{k}}\varphi_{\mathbf{k'}}\sigma_{\mathbf{q}}U(\tau',\tau)~,
\end{align}
which represents the first order term of $G(\tau)$ operating to momentum conserving modes $\mathbf{k,k',q}$, and it satisfies the initial condition
\begin{align}
\mathcal{K}_{\mathbf{k,k',q}}(\tau,\tau)=\varphi_{\mathbf{k}}\varphi_{\mathbf{k'}}\sigma_{\mathbf{q}}~.
\end{align}
Also, this operator can be regarded as a Green's function to connect the state at time $\tau$ to intermediate time $\tau'$. For showing the dependence of this quantity on $\tau'$, it is suitable to find its differential equation
\begin{align}\label{eq:differential equation for K}
\frac{\partial\mathcal{K}_{\mathbf{k,k',q}}(\tau',\tau)}{\partial\tau'}=ia(\tau)U^{-1}(\tau',\tau)[H_{0}[\varphi(\tau_{0}),\sigma(\tau_{0}),\tau'],\varphi_{\mathbf{k}}\varphi_{\mathbf{k'}}\sigma_{\mathbf{q}}]U(\tau',\tau)~.
\end{align}
With Equation (\ref{eq:Hamiltonain}) and canonical commutation relation, we can work out the commutator in Equation (\ref{eq:differential equation for K}) after some algebras,
\begin{align}\label{eq:commutator}
[H_{0}[\varphi(\tau_{0}),\sigma(\tau_{0}),\tau'],\varphi_{\mathbf{k}}\varphi_{\mathbf{k'}}\sigma_{\mathbf{q}}]=-if(\tau')\Big(\pi_{\mathbf{-k}}^{(\varphi)}\varphi_{\mathbf{k'}}\sigma_{\mathbf{q}}+\varphi_{\mathbf{k}}\pi_{\mathbf{-k}'}^{(\varphi)}\sigma_{\mathbf{q}}+\varphi_{\mathbf{k}}\varphi_{\mathbf{k'}}\pi_{\mathbf{-q}}^{(\sigma)}\Big)~,
\end{align}
where
\begin{align}
f(\tau')=\frac{1}{a(\tau')^{3}}~.
\end{align}
The Gaussian state $|\psi_{\text{G}}(\tau)\rangle$ does not depend on the intermediate time, and thus the intermediate time derivative of the Green's function operating on Gaussian wave functional (source) can be evaluated as
\begin{align}\label{eq:differential equation of K bracket}
&\frac{\partial\langle\varphi,\sigma|\mathcal{K}_{\mathbf{k,k',q}}(\tau',\tau)|\psi_{\text{G}}(\tau)\rangle}{\partial\tau'}\nonumber\\
&=ia(\tau')f(\tau')\Big(A_{\varphi}(k,\tau')+A_{\varphi}(k',\tau')+A_{\sigma}(q,\tau')\Big)\langle\varphi,\sigma|\mathcal{K}_{\mathbf{k,k',q}}(\tau',\tau)|\psi_{\text{G}}(\tau)\rangle~,
\end{align}
With the initial condition of $\mathcal{K}_{\mathbf{k,k',q}}(\tau',\tau)$, the c-number functional $\langle\varphi,\sigma|\mathcal{K}_{\mathbf{k,k',q}}(\tau',\tau)|\psi_{G}(\tau)\rangle$ can be solved, and it is expressed in exponential form
\begin{align}
\langle\varphi,\sigma|\mathcal{K}_{\mathbf{k,k',q}}(\tau',\tau)|\psi_{G}(\tau)\rangle=\exp\left({i\alpha(\tau',\tau)}\right)\varphi_{\mathbf{k}}\varphi_{\mathbf{k'}}\sigma_{\mathbf{q}}\langle\varphi,\sigma|\psi_{G}(\tau)\rangle\label{eq:solve K bracket}~,
\end{align}
where $\varphi_{\mathbf{k}}\varphi_{\mathbf{k'}}\sigma_{\mathbf{q}}$ is the eigenvalue of field-amplitude basis, and
\begin{align}
\exp\left({i\alpha(\tau',\tau)}\right)=\exp\left({-i\int_{\tau'}^{\tau}d\tau''a(\tau'')f(\tau'')\Big(A_{\varphi}(k,\tau'')+A_{\varphi}(k',\tau'')+A_{\sigma}(q,\tau'')\Big)}\right)~.
\end{align}
After operating the Green's function on the Gaussian wave functional,
extra factor appears. With weak enough coupling, the exponential
of this extra factor can approximate the factor generated by $G(\tau)$
with Equation (\ref{eq:G from interaction picture}). Therefore, the total
wave functional is the product of the Gaussian part and the non-Gaussian part,
\begin{align}\label{eq:total wave functional}
\Psi[\varphi,\sigma]=\exp\left(\int_{\mathbf{k,k',q}}\mathcal{F}_{\mathbf{k,k',q}}(\tau)\varphi_{\mathbf{k}}\varphi_{\mathbf{k'}}\sigma_{\mathbf{q}}\right)\Psi_{\text{G}}[\varphi,\sigma]~,
\end{align}
where
\begin{align}
&\mathcal{F}_{\mathbf{k,k',q}}(\tau)=-i\int_{\tau_{0}}^{\tau}d\tau'\ a(\tau')\tilde{\mathcal{H}}_{\mathbf{k,k',q}}^{(\text{int})}(\tau')\times\nonumber\\
&\exp\left({-i\int_{\tau'}^{\tau}d\tau''a(\tau'')f(\tau'')\Big(A_{\varphi}(k,\tau'')+A_{\varphi}(k',\tau'')+A_{\sigma}(q,\tau'')\Big)}\right)~.
\end{align}
The integrand consists of a scale factor, interaction in momentum space and a factor generated by the Green function which does not depend on coupling strength. With the exact solution given by free field theory, the factor generated by the Green's function has explicit form that is the ratio of the product of three mode functions at intermediate time $\tau'$ to final conformal time $\tau$,
\begin{align}\label{eq:three Hankel}
&\exp\left({-i\int_{\tau'}^{\tau}d\tau''a(\tau'')f(\tau'')\Big(A_{\varphi}(k,\tau'')+A_{\varphi}(k',\tau'')+A_{\sigma}(q,\tau'')\Big)}\right)\nonumber\\
=&\exp\left(\int_{\tau}^{\tau'}d\tau''\ \frac{\partial\Big(\log(u_{k}(\tau'')u_{k'}(\tau'')w_{q}(\tau''))\Big)}{\partial\tau''}\right)=\frac{u_{k}(\tau')u_{k'}(\tau')w_{q}(\tau')}{u_{k}(\tau)u_{k'}(\tau)w_{q}(\tau)}~.
\end{align}
Combining the interaction and these mode functions, the exponent for non-Gaussian part of wave functional can be expressed by an integral
\begin{align}
\mathcal{F}_{\mathbf{k,k',q}}(\tau)=\frac{ig(q^{2}-k^{2}-k'^{2})}{2H^{2}}\frac{\int_{\tau_{0}}^{\tau}d\tau'(-\tau')^{\frac{5}{2}}H_{\frac{3}{2}}^{(2)}(-k\tau')H_{\frac{3}{2}}^{(2)}(-k'\tau')H_{\nu}^{(2)}(-q\tau')}{(-\tau)^{\frac{9}{2}}H_{\frac{3}{2}}^{(2)}(-k\tau)H_{\frac{3}{2}}^{(2)}(-k'\tau)H_{\nu}^{(2)}(-q\tau)}~.\label{eq:F}
\end{align}
Conceptually, the conformal time integral in Equation (\ref{eq:F}) should be evaluated after the momentum integral in Equation (\ref{eq:total wave functional}) because the non-Gaussian part is generated by the time integral of the interaction Hamiltonian, as shown in Equation (\ref{eq:G from interaction picture}). We assume that the order of integration is interchangeable at the moment, and the advantage of defining $\mathcal{F}_{\mathbf{k,k',q}}(\tau)$ and writing the momentum integral explicitly is easier to trace out the environment. Together with Equation (\ref{eq:total wave functional}), the non-Gaussian part involves every momentum conserving modes at different conformal time. The wave functional of the state consists of inflaton and massive scalar field is calculated in interaction theory, and the next step is to determine the reduced density matrix.

\subsection{Reduced density matrix and decoherence rate}

As mentioned in subsection \ref{decoherence theory}, the decoherence is described by the vanishing off-diagonal terms of the reduced density matrix which is defined by tracing out the whole density matrix with environment. In the case of field theory, this matrix can be represented by a functional that takes two field configurations as arguments,
\begin{align}\label{eq:reduced density matrix}
\rho_{R}[\sigma,\tilde{\sigma}]&=\Psi_{\text{G}}^{(\sigma)}[\sigma](\Psi_{\text{G}}^{(\sigma)}[\tilde{\sigma}])^{*}\int\mathcal{D}\varphi|\Psi_{\text{G}}^{(\varphi)}|^{2}\exp\Big[\int_{\mathbf{k,k',q}}\varphi_{\mathbf{k}}\varphi_{\mathbf{k'}}(\sigma_{\mathbf{q}}\mathcal{F}_{\mathbf{k,k',q}}+\tilde{\sigma}_{\mathbf{q}}\mathcal{F}_{\mathbf{k,k',q}}^{*})\Big]\nonumber\\
&=\Psi_{\text{G}}^{(\sigma)}[\sigma](\Psi_{\text{G}}^{(\sigma)}[\tilde{\sigma}])^{*}\Big(1+\langle X\rangle_{\varphi}+\frac{\langle X^{2}\rangle_{\varphi}}{2!}+\mathcal{O}(\mathcal{F}^{3})\Big)\nonumber\\
&=\Psi_{\text{G}}^{(\sigma)}[\sigma](\Psi_{\text{G}}^{(\sigma)}[\tilde{\sigma}])^{*}\Big(1+\langle X\rangle_{\varphi}+\frac{\langle X\rangle_{\varphi}^{2}}{2!}+\mathcal{O}(\mathcal{F}^{3})\Big)\Big(1+\frac{\langle X^{2}\rangle_{\varphi}-\langle X\rangle_{\varphi}^{2}}{2!}+\mathcal{O}(\mathcal{F}^{3})\Big)\nonumber\\
&\approx\Psi_{\text{G}}^{(\sigma)}[\sigma](\Psi_{\text{G}}^{(\sigma)}[\tilde{\sigma}])^{*}\exp\left({\langle X\rangle_{\varphi}+\frac{\mathrm{Var}{}_{\varphi}[X]}{2}}\right)~,
\end{align}
where the variable $X$ is defined as
\begin{align}
X&=\int_{\mathbf{k,k',q}}\varphi_{\mathbf{k}}\varphi_{\mathbf{k'}}(\sigma_{\mathbf{q}}\mathcal{F}_{\mathbf{k,k',q}}+\tilde{\sigma}_{\mathbf{q}}\mathcal{F}_{\mathbf{k,k',q}}^{*})\nonumber\\
&=\int_{\mathbf{k,k',q}}\varphi_{\mathbf{k}}\varphi_{\mathbf{k'}}\Big[(\sum\sigma_{\mathbf{q}})\mathrm{Re}\mathcal{F}_{\mathbf{k,k',q}}+i(\Delta\sigma_{\mathbf{q}})\mathrm{Im}\mathcal{F}_{\mathbf{k,k',q}}\Big]~,
\end{align}
where $\Sigma$ means summation of two fields, while $\Delta$ means difference of two fields, and the functional integration,
\begin{align}
\int\mathcal{D}\varphi|\Psi_{\text{G}}^{(\varphi)}|^{2}\left(...\right)=\langle...\rangle_\varphi~,
\end{align}
and the statistical variance,
\begin{align}
\text{Var}_\varphi[X]={\langle X^{2}\rangle_{\varphi}-\langle X\rangle_{\varphi}^{2}}~.
\end{align}
Clearly, the reduced density matrix has the form that is proportional to the moment generating function of Gaussian distribution with unitary parameter. To avoid the difficulty of normalizing the wave functional, the decay of off-diagonal terms can be expressed by the absolute value of the ratio of off-diagonal terms to diagonal terms. One important thing is that in \cite{Nelson:2016kjm} the real part of $\mathcal{F}_{\mathbf{k,k',q}}(\tau)$ is supposed to
be small compared to its imaginary part, but this may be important when we integrate high energy modes, and therefore it is kept here. We consider the expectation value of $X$ in Equation (\ref{eq:reduced density matrix}) first. After taking absolute value to the ratio, the term that is proportional to $\mathrm{Re}\mathcal{F}_{\mathbf{k,k',q}}$ in $X$ is cancelled, and another term in $X$ remains imaginary with the symmetry
\begin{align}
\langle\varphi_{\mathbf{k}}\varphi_{\mathbf{k'}}\rangle_{\varphi}(\Delta\sigma_{-\mathbf{q}})\mathrm{Im}\mathcal{F}_{\mathbf{k,k',-q}}=\langle\varphi_{\mathbf{k}}\varphi_{\mathbf{k'}}\rangle_{\varphi}(\Delta\sigma_{\mathbf{q}})^{*}\text{Im}\mathcal{F}_{\mathbf{k,k',q}}~,
\end{align}
so the term $\exp\left({\langle X\rangle_{\varphi}}\right)$ contributes nothing to the result. Thus, it depends only on the variance of $X$, and we define $X_{1}$ and $X_{2}$ are the integral whose massive field arguments are all $\sigma_{\mathbf{q}}$ and $\tilde{\sigma}_{\mathbf{q}}$ respectively, then
\begin{align}\label{eq:D matrix}
D[\sigma,\tilde{\sigma}]=\Big|\frac{\rho_{R}[\sigma,\tilde{\sigma}]}{\sqrt{\rho_{R}[\sigma,\sigma]\rho_{R}[\tilde{\sigma},\tilde{\sigma}]}}\Big|\approx\Big|\text{exp}\Big[\frac{1}{2}\Big(\mathrm{Var}{}_{\varphi}[X]-\frac{\mathrm{Var}{}_{\varphi}[X_{1}]+\mathrm{Var}{}_{\varphi}[X_{2}]}{2}\Big)\Big]\Big|~.
\end{align}
The variance is related to the power spectrum of inflaton by
\begin{align}
\langle\varphi_{\mathbf{k}}\varphi_{\mathbf{k'}}\rangle_{\varphi}=(2\pi)^{3}\delta^{3}(\mathbf{k}+\mathbf{k'})P_{\varphi}(k,\tau)=\frac{(2\pi)^{3}\delta^{3}(\mathbf{k}+\mathbf{k'})}{2\mathrm{Re}\{A_{\varphi}(k,\tau)\}}~.
\end{align}
Then one can show that
\begin{align}\label{eq:Var}
\text{Var}_\varphi[X]&=\int_{\mathbf{k}_{1},\mathbf{k}_{1}',\mathbf{q}_{1}}\int_{\mathbf{k}_{2},\mathbf{k}_{2}',\mathbf{q}_{2}}\Big(\langle\varphi_{\mathbf{k}_1}\varphi_{\mathbf{k}_2}\rangle_{\varphi}\langle\varphi_{\mathbf{k}_1'}\varphi_{\mathbf{k}_2'}\rangle_{\varphi}+(\mathbf{k}_{2}\rightarrow\mathbf{k}_{2}')\Big)\nonumber\\
&\times\Big[(\sigma_{\mathbf{q}_1}\mathcal{F}_{\mathbf{k}_{1},\mathbf{k}_{1}',\mathbf{q}_{1}}+\tilde{\sigma}_{\mathbf{q}_{1}}\mathcal{F}_{\mathbf{k}_{1},\mathbf{k}_{1}',\mathbf{q}_{1}}^{*{\color{black}}})(1\rightarrow2)\Big]\nonumber\\
&=2\int_{\mathbf{k,k',q}}P_{\varphi}(k,\tau)P_{\varphi}(k',\tau)\nonumber\\
&\times\Big\{\Big[(\sum\sigma_{\mathbf{q}})\mathrm{Re}\mathcal{F}_{\mathbf{k,k',q}}+i(\Delta\sigma_{\mathbf{q}})\mathrm{Im}\mathcal{F}_{\mathbf{k,k',q}}\Big](\mathbf{q}\to-\mathbf{q})\Big\}~.
\end{align}
Then the exponent in Equation (\ref{eq:D matrix}) can be evaluated easily as
\begin{align}\label{eq:exponent}
&\frac{1}{2}\Big(\mathrm{Var}{}_{\varphi}[X]-\frac{\mathrm{Var}{}_{\varphi}[X_{1}]+\mathrm{Var}{}_{\varphi}[X_{2}]}{2}\Big)=\int_{\mathbf{k,k',q}}P_{\varphi}(k,\tau)P_{\varphi}(k',\tau)\times\nonumber\\
&\Big[-|\Delta\sigma_{\mathbf{q}}|^{2}\Big|\mathcal{F}_{\mathbf{k,k',q}}\Big|^{2}+i(\mathrm{Re}\mathcal{F}_{\mathbf{k,k',q}})(\mathrm{Im}\mathcal{F}_{\mathbf{k,k',q}})(\left(\sum\sigma_{\mathbf{q}}\right)\left(\Delta\sigma_{\mathbf{q}}\right)+\text{c.c.})\Big]~.
\end{align}
The second term in Equation (\ref{eq:exponent}) disappears after taking absolute value, and eventually the quantity depends on the absolute square of $\mathcal{F}_{\mathbf{k,k',q}}(\tau)$. During inflationary epoch, the subhorizon and superhorizon modes behaves differently, and it is proper to study the reduced density matrix for a mode with momentum $\mathbf{q}$. To get rid of the integral with respect to $\mathbf{q}$, we use
\begin{align}
\int\frac{d^{3}\mathbf{q}}{(2\pi)^{3}}=\frac{\sum_{\mathbf{q}}}{(2\pi)^{3}\delta^{3}(0)}=\frac{\sum_{\mathbf{q}}}{V}~.
\end{align}
Thus we obtain
\begin{align}\label{eq:ratio at q}
D[\sigma,\tilde{\sigma}](\mathbf{q},\tau)=\exp\Big[-\frac{|\Delta\sigma_{\mathbf{q}}|^{2}}{V}\int_{\mathbf{k+k'=-q}}P_{\varphi}(k,\tau)P_{\varphi}(k',\tau)\Big|\mathcal{F}_{\mathbf{k,k',q}}\Big|^{2}\Big]~,
\end{align}
where the symbol $\int_{\mathbf{k+k'=-q}}=\int\frac{d^{3}\mathbf{k}}{(2\pi)^{3}}\frac{d^{3}\mathbf{k'}}{(2\pi)^{3}}(2\pi)^{3}\delta^{3}(\mathbf{k}+\mathbf{k'}+\mathbf{q})$
means integration over the modes of inflaton. For the reduced density matrix in quantum mechanics, it usually satisfies the equation \cite{Zurek3}
\begin{align}
\rho_{R}(x,x',t)=\rho_{R}(x,x',0)\exp\left({-\gamma t\left(\frac{x-x'}{\lambda_T}\right)^{2}}\right)~,
\end{align}
where $\gamma$ represents the relaxation coefficient and $\lambda_T$ is the thermal de Broglie wavelength. And thus the absolute value of ratio of off-diagonal terms to diagonal terms is
\begin{align}
\left|\frac{\rho_{R}(x,x',t)}{\sqrt{\rho_{R}(x,x,t)\rho_{R}(x',x',t)}}\right|&=\left|\frac{\rho_{R}(x,x',0)}{\sqrt{\rho_{R}(x,x,0)\rho_{R}(x',x',0)}}\exp\left({-\gamma t\left(\frac{x-x'}{\lambda_T}\right)^{2}}\right)\right|\nonumber\\
&=\left|\exp\left({-\gamma t\left(\frac{x-x'}{\lambda_T}\right)^{2}}\right)\right|~.
\end{align}
The ratio at $t=0$ is one because the initial pure state have linear combination form given by Equation (\ref{eq:density matrix}). Moreover, a decoherence rate can be defined by the time scale when the exponent becomes $\mathcal{O}(1)$. Similarly, Equation (\ref{eq:ratio at q}) describes the extent of decoherence for massive field's mode with momentum $\mathbf{q}$, and the dimensionless decoherence rate can also be defined as $-\mathrm{log}\Big(D[\sigma,\tilde{\sigma}](\mathbf{q},\tau)\Big)$, so the corresponding dimensionless time-scale can be defined as the e-folds after crossing horizon, namely $-\log(-q\tau)$ which is called delay in \cite{Nelson:2016kjm}. Combining Equation (\ref{eq:F}) and Equation (\ref{eq:ratio at q}), the decoherence rate is expressed as one-loop integration with respect to momentum space and conformal time,
\begin{align}\label{eq:decoherence rate}
&\Gamma_\text{deco}(\mathbf{q},\tau)=\frac{|\Delta\sigma_{\mathbf{q}}|^{2}g^{2}}{16V(-\tau)^{3}\Big|H_{\nu}^{(2)}(-q\tau)\Big|^{2}}\int_{\tau_{0}}^{\tau}d\tau'\int_{\tau_{0}}^{\tau}d\tau''(\tau'\tau'')^{-\frac{1}{2}}H_{\nu}^{(2)}(-q\tau')H_{{\nu}^*}^{(1)}(-q\tau'')\nonumber\\
&\times\int_{\mathbf{k+k'=-q}}\frac{(q^{2}-k^{2}-k'^{2})^{2}e^{i(k+k')(\tau'-\tau'')}(i+k\tau')(i+k'\tau')(-i+k\tau'')(-i+k'\tau'')}{k^{3}k'^{3}}~.
\end{align}
The integral over momentum space means summing over all the modes of the environment which consists of the inflaton, and therefore the integrand is quite simple.

\subsection{One-loop computation}

\begin{figure}[htbp]
  \centering
  \includegraphics[width=0.6\textwidth]{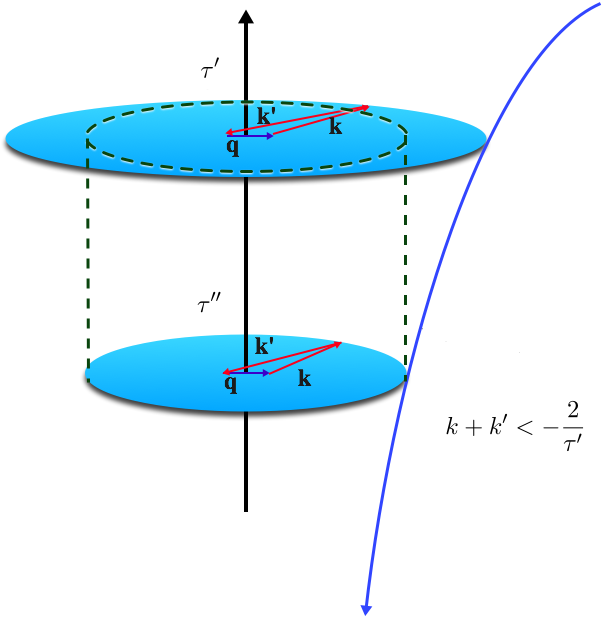}
  \caption{\label{corre} The UV-cutoff at different time. The region inside the cylinder can render all the modes are superhorizon at given conformal time.}
\end{figure}

\begin{figure}[htbp]
  \centering
  \includegraphics[width=0.6\textwidth]{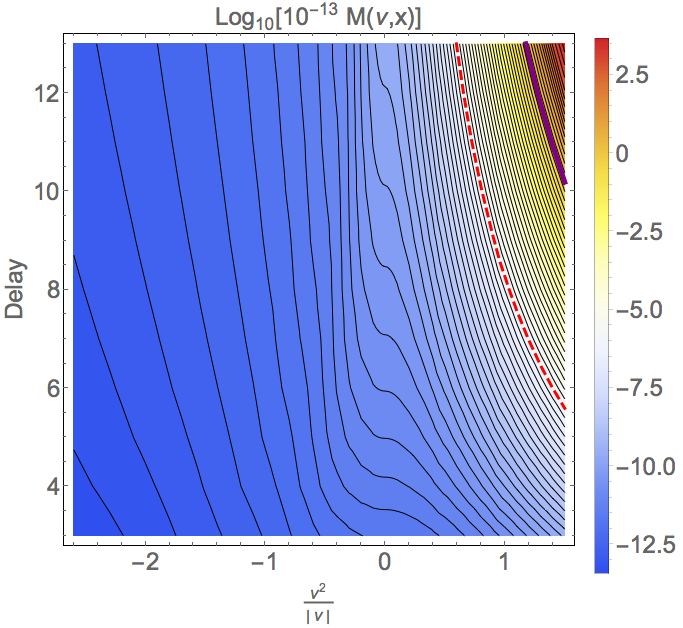}
  \caption{\label{numer} The red dash line is for the values with $\mathcal{O}(10^{-6})$ which corresponds to $\frac{(gH)^{2}}{8\pi^{2}}=10^{-7}$, whereas the solid purple line is for the value with $\mathcal{O}(1)$ which corresponds to $\frac{(gH)^{2}}{8\pi^{2}}=10^{-13}$. The difference of two adjoin
contours is $0.25$ that means the ratio of decoherence rate is $10^{0.25}$.}
\end{figure}

From the integrand of the momentum part, the exponent of the oscillating term is proportional to $(\tau'-\tau'')$, and we introduce a new coordinate
\begin{align}
\mathcal{U}=q(\tau'-\tau'')~~~,~~~\mathcal{V}=q(\tau'+\tau'')~,
\end{align}
and the Jacobian is $2q^{2}$. The integral over momentum space can be evaluated with elliptical coordinate which introduces variables \cite{Jiang 2016}
\begin{align}
\mu_1=\frac{k+k'}{q}~~~,~~~\mu_2=\frac{k-k'}{q}~.
\end{align}
Also, we define the variable $x=-q\tau$ and $x_0=-q\tau_0$, then the decoherence rate is given as
\begin{align}
{{\Gamma }_{\text{deco}}}(x)=\frac{|\Delta {{\sigma }_{\mathbf{q}}}{{|}^{2}}{{g}^{2}}{{q}^{3}}\left( \int_{-{{x}_{0}}-x}^{-2x}{d}\mathcal{V}\int_{\mathcal{V}+2x}^{-2x-\mathcal{V}}{d}\mathcal{U}+\int_{-2{{x}_{0}}}^{-{{x}_{0}}-x}{d}\mathcal{V}\int_{-2{{x}_{0}}-\mathcal{V}}^{\mathcal{V}+2{{x}_{0}}}{d}\mathcal{U} \right)\mathcal{R} (\mathcal{U},\mathcal{V})}{8{{\pi }^{2}}V{{x}^{3}}|H_{\nu }^{(2)}(x){{|}^{2}}}~,
\end{align}
where
\begin{align}
  & \mathcal{R} (\mathcal{U},\mathcal{V})={{({{\mathcal{V}}^{2}}-{{\mathcal{U}}^{2}})}^{-\frac{1}{2}}}H_{\nu }^{(2)}(-\frac{\mathcal{V}+\mathcal{U}}{2})H_{{{\nu }^{*}}}^{(1)}(-\frac{\mathcal{V}-\mathcal{U}}{2})\int_{1}^{+\infty }{d}{{\mu }_{1}}\int_{-1}^{1}{d}{{\mu }_{2}}\ \times   \nonumber\\
 & {{\left( 1-\frac{\mu _{1}^{2}+\mu _{2}^{2}}{2} \right)}^{2}}\frac{{{e}^{i{{\mu }_{1}}\mathcal{U}}}}{{{(\mu _{1}^{2}-\mu _{2}^{2})}^{2}}}\left( i+\frac{({{\mu }_{1}}+{{\mu }_{2}})(\mathcal{V}+\mathcal{U})}{4} \right)\left( i+\frac{({{\mu }_{1}}-{{\mu }_{2}})(\mathcal{V}+\mathcal{U})}{4} \right)\times  \nonumber\\
 & \left( -i+\frac{({{\mu }_{1}}+{{\mu }_{2}})(\mathcal{V}-\mathcal{U})}{4} \right)\left( -i+\frac{({{\mu }_{1}}-{{\mu }_{2}})(\mathcal{V}-\mathcal{U})}{4} \right)~.
\end{align}
The integrand of momentum space in $\mathcal{R}$ is proportional to $e^{i\mu_1\mathcal{U}}\mu_1^{4}$ when $\mu_1\to+\infty$, and thus the modes with squeezed limit
\begin{align}
\mu_1=\frac{k+k'}{q}\gg1~,
\end{align}
make main contributions. The integration over $\mu_{1,2}$ can be written down as
\begin{align}
\mathcal{R} (\mathcal{U},\mathcal{V})={{\mathcal{J}}_{1}}(\mathcal{U},\mathcal{V}){{\mathcal{J}}_{2}}(\mathcal{U},\mathcal{V})~,
\end{align}
where
\begin{align}
 {{\mathcal{J}}_{1}}(\mathcal{U},\mathcal{V})&= {{({{\mathcal{V}}^{2}}-{{\mathcal{U}}^{2}})}^{-\frac{1}{2}}}H_{\nu }^{(2)}(-\frac{\mathcal{V}+\mathcal{U}}{2})H_{{{\nu }^{*}}}^{(1)}(-\frac{\mathcal{V}-\mathcal{U}}{2})e^{i\mathcal{U}} ~,\nonumber \\
  {{\mathcal{J}}_{2}}(\mathcal{U},\mathcal{V})&=\frac{3i{{\mathcal{V}}^{4}}}{64{{\mathcal{U}}^{5}}}+\frac{3{{\mathcal{V}}^{4}}}{64{{\mathcal{U}}^{4}}}-\frac{i\left( {{\mathcal{V}}^{4}}+39{{\mathcal{V}}^{2}} \right)}{96{{\mathcal{U}}^{3}}}+\frac{{{\mathcal{V}}^{4}}-78{{\mathcal{V}}^{2}}}{192{{\mathcal{U}}^{2}}}+\frac{i\left( {{\mathcal{V}}^{4}}-60{{\mathcal{V}}^{2}}+1545 \right)}{960\mathcal{U}}  \nonumber \\
 & +\frac{{{\mathcal{V}}^{2}}}{96}+\frac{7}{64}-\frac{i\mathcal{U}\left( {{\mathcal{V}}^{2}}+5 \right)}{480}-\frac{{{\mathcal{U}}^{2}}}{64}+\frac{i{{\mathcal{U}}^{3}}}{960} ~.
\end{align}
There is high order pole at $\mathcal{U}=0$. When $\mathcal{V}\to-2x$ or $-2x_{0}$, the whole contour is closed to the high order pole. Expanding the integrand near $\mathcal{U}=0$, the integral has form
\begin{align}
\int_{-\epsilon}^{\epsilon}d\mathcal{U}\sum_{n=-5}^{\infty}a_{n}(\mathcal{V})\mathcal{U}^{n}~,
\end{align}
and therefore the divergent term
\begin{align}
\sum_{n=1}^{4}\frac{b_{n}(\mathcal{V})}{\epsilon^{n}}~,
\end{align}
appears. The divergence is attributed to zero exponent of oscillating factor $e^{i\mu_1\mathcal{U}}$, and the integrand is reduced to rational function with asymptotic form $\mu_1^4$, so divergence $\mu_{1}^{5}$ appears. The physical meaning is that this one-loop is UV-divergent, thus we need to make a cutoff.
\\
\\
Instead of studying decoherence for all possible modes, we can simply focus on superhorizon modes. One may simply choose the horizon at the end of interaction as a cutoff,
\begin{align}
k_{\text{max}}=-\frac{1}{\tau}~,
\end{align}
but then the integrand of momentum integral evaluated at initial time may be very big since
\begin{align}
-k_{\text{max}}\tau_{0}=\frac{\tau_{0}}{\tau}=\frac{a(\tau)}{a(\tau_{0})}~.
\end{align}
The physical meaning is that such a cutoff includes many superhorizon modes at early time, and it is sure that the integral cannot converge. On the other hand, if the UV-cutoff was selected as the superhorizon at the initial time, many contributions would have been neglected. However, we can choose a time-dependent cutoff which makes the massive field to interact superhorizon modes of inflaton only, and clearly the integral converges. Mathematically, the conditions for environment are assumed as
\begin{align}
-k\tau',-k'\tau'<\mathcal{O}(1)~,
\end{align}
that means the related inflaton modes are superhorizon at every intermediate time $\tau'$. With the triangular inequality for the three interacting modes
\begin{align}
-q\tau'\leq-(k+k')\tau'<\mathcal{O}(1)~,
\end{align}
the modes of massive field are also required to be superhorizon during interaction, and this constraints the initial time for interaction. A theory satisfying
these constraints describes the interaction between superhorizon modes, and therefore our model becomes an effective theory to describe the decoherence due to the interactions between superhorizon modes. The next step is to modify the interval of integration in Equation (\ref{eq:decoherence rate}). The number of momentum integral is one, but the integrand involves two conformal time $\tau',\tau''$ which have different UV-cutoffs
$\frac{2}{-q\tau'}$ and $\frac{2}{-q\tau''}$ for $\mu_1$, so the choice of UV-cutoff is nontrivial. Intuitively, the minimum value $\mathrm{Min}\{\frac{2}{-q\tau'} ,\frac{2}{-q\tau''}\}$ which corresponds to all superhorizon modes at $\mathrm{Min}\{\tau',\tau''\}$ should be selected for ensuring all the terms $k\tau',k\tau''\leq1$, as shown in the Fig.~\ref{corre}.
Mathematically, this can be proved by reconsidering Equation (\ref{eq:Var}) with time dependent UV-cutoff. We define a practical $\mathcal{F}$ in our analytic investigations as
\begin{align}
&\mathcal{{F}}_{\mathbf{k,k',q}}(\tau')=-ia(\tau')\tilde{\mathcal{H}}_{\mathbf{k,k',q}}^{(\text{int})}(\tau')\nonumber\\
&\exp\left({-i\int_{\tau'}^{\tau}d\tau''a(\tau'')f(\tau'')\Big(A_{\varphi}(k,\tau'')+A_{\varphi}(k',\tau'')+A_{\sigma}(q,\tau'')\Big)}\right)~,
\end{align}
in order to put the time integral to the front. From now on, the interchangeability of time and momentum integral is broken because of the time dependent cutoff. The two point correlation function in momentum space is non-zero only if total momentum of the modes are zero, and therefore the non-vanish correlations are with the smaller cutoff,
\begin{align}
   &\text{Va}{{\text{r}}_{\varphi }}[X]=\int_{{{\tau }_{0}}}^{\tau }{d{\tau }'\int_{{{\tau }_{0}}}^{\tau }{d{\tau }''}}\int_{{{\mathbf{k}}_{1}},{{{\mathbf{{k}}}}_{1}'},{{\mathbf{q}}_{1}}}^{<\frac{1}{-{\tau }'}}{\int_{{{\mathbf{k}}_{2}},{{{\mathbf{{k}}}}_{2}'},{{\mathbf{q}}_{2}}}^{<\frac{1}{-{\tau }''}}{\left( {{\left\langle {{\varphi }_{{{\mathbf{k}}_{1}}}}{{\varphi }_{{{\mathbf{k}}_{2}}}} \right\rangle }_{\varphi }}{{\left\langle {{\varphi }_{{{{\mathbf{{k}}}}_{1}'}}}{{\varphi }_{{{{\mathbf{{k}}}}_{2}'}}} \right\rangle }_{\varphi }}+({{\mathbf{k}}_{2}}\to {{{\mathbf{{k}}}}_{2}'}) \right)}} \nonumber\\
 & \times \left( ({{\sigma }_{{{\mathbf{q}}_{1}}}}{{\mathcal{F}}_{{{\mathbf{k}}_{1}},{{{\mathbf{{k}}}}_{1}'},{{\mathbf{q}}_{1}}}}({\tau }')+{{{\tilde{\sigma }}}_{{{\mathbf{q}}_{1}}}}\mathcal{F}_{{{\mathbf{k}}_{1}},{{{\mathbf{{k}}}}_{1}'},{{\mathbf{q}}_{1}}}^{*}({\tau }'))(1\to 2,{\tau }'\to {\tau }'') \right) \nonumber\\
 &=\int_{{{\tau }_{0}}}^{\tau }{d{\tau }'\int_{{{\tau }_{0}}}^{\tau }{d{\tau }''}}\int_{{{\mathbf{k}}_{1}},{{{\mathbf{{k}}}}_{1}'},{{\mathbf{q}}_{1}}}^{<\text{Min}\left\{ \frac{1}{-{\tau }'},\frac{1}{-{\tau }''} \right\}}{\int_{{{\mathbf{k}}_{2}},{{{\mathbf{{k}}}}_{2}'},{{\mathbf{q}}_{2}}}^{<\text{Min}\left\{ \frac{1}{-{\tau }'},\frac{1}{-{\tau }''} \right\}}{\left( {{\left\langle {{\varphi }_{{{\mathbf{k}}_{1}}}}{{\varphi }_{{{\mathbf{k}}_{2}}}} \right\rangle }_{\varphi }}{{\left\langle {{\varphi }_{{{{\mathbf{{k}}}}_{1}'}}}{{\varphi }_{{{{\mathbf{{k}}}}_{2}'}}} \right\rangle }_{\varphi }}+({{\mathbf{k}}_{2}}\to {{{\mathbf{{k}}}}_{2}'}) \right)}} \nonumber\\
  &\times \left( ({{\sigma }_{{{\mathbf{q}}_{1}}}}{{\mathcal{F}}_{{{\mathbf{k}}_{1}},{{{\mathbf{{k}}}}_{1}'},{{\mathbf{q}}_{1}}}}({\tau }')+{{{\tilde{\sigma }}}_{{{\mathbf{q}}_{1}}}}\mathcal{F}_{{{\mathbf{k}}_{1}},{{{\mathbf{{k}}}}_{1}'},{{\mathbf{q}}_{1}}}^{*}({\tau }'))(1\to 2,{\tau }'\to {\tau }'') \right)~.
\end{align}
For two distant conformal times, lots of superhorizon modes at the later time are lost with the smaller cutoff, and the rest remains low energy. This leads to smaller contribution to the integral, compared to the case with two adjoin conformal times. A rescaled conformal time coordinate
\begin{align}
x'=-q\tau'~~~,~~~ x''=-q\tau''~,
\end{align}
is introduced to show the time dependent cutoff explicitly in the expression of decoherence rate. To work out the integral numerically, the cutoff is represented by the Heaviside function $\Theta$ as
\begin{align}
  & {{\Gamma }_{\text{deco}}}(x)=\frac{|\Delta {{\sigma }_{\mathbf{q}}}{{|}^{2}}{{g}^{2}}{{q}^{3}}}{8{{\pi }^{2}}V{{x}^{3}}|H_{\nu }^{(2)}(x){{|}^{2}}}\left( \int_{x}^{2}{d}{x}'\int_{x}^{2}{d}{x}'' \right){{\left( {x}'{x}'' \right)}^{-\frac{1}{2}}}H_{\nu }^{(2)}({x}')H_{{{\nu }^{*}}}^{(1)}({x}'') \nonumber\\
 & \int_{1}^{\frac{2}{x}}{d}{{\mu }_{1}}\int_{-1}^{1}{d{{\mu }_{2}}}\ {{\left( 1-\frac{\mu _{1}^{2}+\mu _{2}^{2}}{2} \right)}^{2}}\Theta \left( \text{Min}\left\{ \frac{2}{{{x}'}},\frac{2}{{{x}''}} \right\}-{{\mu }_{1}} \right)\frac{{{e}^{i{{\mu }_{1}}({x}''-{x}')}}}{{{(\mu _{1}^{2}-\mu _{2}^{2})}^{2}}} \nonumber\\
 & \left( i-\frac{({{\mu }_{1}}+{{\mu }_{2}}){x}'}{2} \right)\left( i-\frac{({{\mu }_{1}}-{{\mu }_{2}}){x}'}{2} \right)\left( -i-\frac{({{\mu }_{1}}+{{\mu }_{2}}){x}''}{2} \right)\left( -i-\frac{({{\mu }_{1}}-{{\mu }_{2}}){x}''}{2} \right) ~.
\end{align}
By interchanging $x'$ and $x''$, the integrand becomes its complex conjugate. We can apply this property to reduce half of the region of integration,
\begin{align}
{{\Gamma }_{\text{deco}}}(x)=\frac{|\Delta {{\sigma }_{\mathbf{q}}}{{|}^{2}}{{g}^{2}}{{q}^{3}}}{8{{\pi }^{2}}V{{x}^{3}}|H_{\nu }^{(2)}(x){{|}^{2}}}2\text{Re}\left( \mathcal{Q} \right)~,
\end{align}
where
\begin{align}\label{eq:Q}
  & \mathcal{Q}=\left( \int_{x}^{2}{d}{x}'\int_{x}^{2}{d}{x}'' \right){{\left( {x}'{x}'' \right)}^{-\frac{1}{2}}}H_{\nu }^{(2)}({x}')H_{{{\nu }^{*}}}^{(1)}({x}'')\Theta \left( {x}''-{x}' \right)\times  \nonumber\\
 & \int_{1}^{\frac{2}{x}}{d}{{\mu }_{1}}\int_{-1}^{1}{d{{\mu }_{2}}}\Theta \left( \frac{2}{{{x}''}}-{{\mu }_{1}} \right){\left( 1-\frac{\mu _{1}^{2}+\mu _{2}^{2}}{2} \right)^{2}}\frac{{{e}^{i{{\mu }_{1}}({x}''-{x}')}}}{{{(\mu _{1}^{2}-\mu _{2}^{2})}^{2}}}\times \nonumber\\
 & \left( i-\frac{({{\mu }_{1}}+{{\mu }_{2}}){x}'}{2} \right)\left( i-\frac{({{\mu }_{1}}-{{\mu }_{2}}){x}'}{2} \right)\left( -i-\frac{({{\mu }_{1}}+{{\mu }_{2}}){x}''}{2} \right)\left( -i-\frac{({{\mu }_{1}}-{{\mu }_{2}}){x}''}{2} \right) ~.
\end{align}
The meaning of this formula is fixing a time $x'$ and sum over all the contributions from earlier time $x''\geq x'$, then integrate all such $x'$. Technically, the integral for $\mu_2$ can be calculated analytically, but the rest of the integrations can only be computed numerically. Besides the integral, a representative absolute square of difference $\Big|\Delta\sigma_{\mathbf{q}}\Big|^{2}$ has to be selected, and it is suitable to evaluate the expectation value of this quantity with Gaussian state,
\begin{align}\label{eq:Variance of massive field}
\langle\psi_{\text{G}}||\Delta\sigma_{\mathbf{q}}|^{2}|\psi_{\text{G}}\rangle(\tau)=2\langle\psi_{\text{G}}|\sigma_{\mathbf{q}}{}^{2}|\psi_{\text{G}}\rangle(\tau)=\frac{V}{\mathrm{Re}\{A_{\sigma}(q,\tau)\}}=V\frac{H^{2}}{q^{3}\mathrm{Re}\left(\frac{i}{x^{2}}\frac{w'(x)}{w(x)}\right)}~.
\end{align}
Thus one can write down the practical formula for one-loop decoherence rate,
\begin{align}\label{eq:ready to compute numerically}
{{\Gamma }_{\text{deco}}}(\nu ,x)=\frac{{{H}^{2}}{{g}^{2}}}{8{{\pi }^{2}}}\frac{1}{{{x}^{3}}|H_{\nu }^{(2)}(x){{|}^{2}}}\frac{2\text{Re}\left( \mathcal{Q} \right)}{\text{Re}\left( \frac{i}{{{x}^{2}}}\frac{{w}'(x)}{w(x)} \right)}=\frac{{{H}^{2}}{{g}^{2}}}{8{{\pi }^{2}}}M(\nu ,x)~.
\end{align}
In this formula, $M(\nu,x)$ is defined and the argument $\nu$ is contained to indicate the mass dependence. Before working out the integral numerically, we would like to comment on Equation (\ref{eq:Q}): The oscillating factor $e^{i\mu_1(x''-x')}$ is attributed to the mode function of inflaton, and the phase is proportional to the difference of conformal time. With the time-dependent cutoff, the change of phase is less than $\pi$, and the effect of interference may not be too large. On the hand, the region with $x''\approx x'$ and $\mathrm{Max}\{x',x''\}\to x$ contributes mainly since the phase is nearly unchanged, the product of Hankel functions becomes absolute square and the UV-cutoff is maximum, and this corresponds to late-time and squeezed limit. The physical meaning of the momentum integral is the correlation of the interaction with the environment at different moments, and therefore the contributions from two distant moments are expected to be small.
\\
\\
Then we numerically computate the mass dependence of the massive cosmic decoherence. The result is shown in Fig.~\ref{numer}. According to the effective field theory analysis between non-Gaussianities and coupling constants \cite{AssassiandBaumann2014}, one can choose the range of the constant as
$\frac{(gH)^{2}}{8\pi^{2}}=\mathcal{O}(10^{-13}\sim10^{-7})$. We put $10^{-13}$ in the numerical computation. The
line with $\mathcal{O}(10^{-6})$ corresponds to the delay for $\frac{(gH)^{2}}{8\pi^{2}}=10^{-7}$.
From Fig.~\ref{numer}, the dense contours in the region of $\frac{\nu^2}{|\nu|}>0$ imply the rapid change of decoherence rate with mass and delay, whereas the sparse and nearly vertical contours in the region of $\frac{\nu^2}{|\nu|}<0$ imply the steady change of decoherence rate. Mathematically, it is related to the boundedness of the Hankel functions appeared in Equation (\ref{eq:Q}). The bounded functions lead the decoherence rate to vary no faster than logarithm, whereas the unbounded functions cause the polynomial divergence at late time, as shown in appendix \ref{app:A}. From the contour plot of decoherence rate, the increment is smaller for heavier field with imaginary $\nu$, and the decoherence for very massive fields is expected to be difficult. For instance, the field with $\frac{\nu^{2}}{|\nu|}<-\frac{1}{2}$, the numerical results
show that the increment per e-folds $k_{\nu}<\mathcal{O}(10^{2})$, and the delay estimated from logarithmic dependence $N\geq\mathcal{O}(10^{5}-10^{11})$
 is much larger than the period of inflation. Therefore, it is
unlikely for the field with $m>\mathcal{O}(H)$ to have decoherence.

\section{Conclusion and Outlook}\label{conclu}

In this paper, we have studied the cosmic decoherence of massive field on the inflationary background. Starting with an introduction on the semiclassical behavior of massive field near the Hubble crossing and a simple introduction of quantum decoherence in quantum information theory, we have then calculated the wave functional for the composite of inflaton and massive field in Schr\"odinger picture, and the relation between the state with interactions and the Gaussian state in free theory is expressed by time-evolution operator based on the cubic interaction between inflaton and massive field.
From
the theory of decoherence, the criterion of losing quantum coherence is the disappearance of off-diagonal terms in reduced density matrix, which is obtained by tracing out the density matrix of quantum state with environment. In Schr\"odinger picture of quantum field theory, the trace is calculated by functional integration, and eventually we find that the reduced density matrix is related to the variance of non-Gaussian exponent of the wave functional.
\\
\\
With mathematical analysis, it is simple to show that the integral throughout the whole momentum space is UV-divergent and the theory is valid when the energy is lower than a UV-cutoff. After rewriting the variance in terms of time integral, it is clear that the decoherence rate is related to two point correlation function at two different moments, leading to the nontrivial choice of UV-cutoff. This is different from the QFT in flat spacetime where the UV-cutoff is usually related to some fixed energy scale such as rest mass of particle, and therefore validity of such cutoff is time independent. For inflationary scenario, the natural cutoff is time-dependent Hubble radius which is related to the distance for causal connection, and this cutoff may be related to the accessibility of information.
\\
\\
Our theory gives a lower bound for decoherence of superhorizon modes. The numerical results show that the quantum coherence for massive field with mass smaller than $\mathcal{O}(H)$ is suppressed within $5\sim10$ e-folds after crossing horizon, whereas the decoherence rate increases slowly for the field with $m\gtrsim \mathcal{O}(H)$, leading to the preservation of quantum coherence. Therefore, massive field is a proper tool to study inflationary quantum information problems.
\\
\\
Some related directions are interesting. First, higher loops can be considered. From Equation (\ref{eq:G from interaction picture}), the second term in the expansion is related to two operators $\mathcal{K}_{\mathbf{k,k',q}}(\tau',\tau)$, and its magnitude is proportional to
$\left(\mathcal{F}_{\mathbf{k,k',q}}(\tau)\right)^{2}$. Therefore, it may affect the reduced density matrix through expectation value according to Equation (\ref{eq:reduced density matrix}). We can evaluate the influence of two $\mathcal{K}_{\mathbf{k,k',q}}(\tau',\tau)$ product to the wave functional,
\begin{align}
  & \langle \varphi ,\sigma |{{\mathcal{K}}_{{{\mathbf{k}}_{1}},{{{\mathbf{{k}}}}_{1}'},{{\mathbf{q}}_{1}}}}({\tau }',\tau ){{\mathcal{K}}_{{{\mathbf{k}}_{2}},{{{\mathbf{{k}}}}_{2}'},{{\mathbf{q}}_{2}}}}({\tau }'',\tau )|{{\psi }_{\text{G}}}(\tau )\rangle  \nonumber\\
 & =\langle \varphi ,\sigma |{{\mathcal{K}}_{{{\mathbf{k}}_{1}},{{{\mathbf{{k}}}}_{1}'},{{\mathbf{q}}_{1}}}}({\tau }',\tau )\int{\mathcal{D}}\tilde{\varphi }\mathcal{D}\tilde{\sigma }|\tilde{\varphi },\tilde{\sigma }\rangle \langle \tilde{\varphi },\tilde{\sigma }|{{\mathcal{K}}_{{{\mathbf{k}}_{2}},{{{\mathbf{{k}}}}_{2}'},{{\mathbf{q}}_{2}}}}({\tau }'',\tau )|{{\psi }_{\text{G}}}(\tau )\rangle   \nonumber\\
 & ={{e}^{i\alpha ({\tau }'',\tau )}}\langle \varphi ,\sigma |{{\mathcal{K}}_{{{\mathbf{k}}_{1}},{{{\mathbf{{k}}}}_{1}'},{{\mathbf{q}}_{1}}}}({\tau }',\tau ){{\varphi }_{{{\mathbf{k}}_{2}}}}{{\varphi }_{{{{\mathbf{{k}}}}_{2}'}}}{{\sigma }_{{{\mathbf{q}}_{2}}}}|{{\psi }_{\text{G}}}(\tau )\rangle   \nonumber\\
 & ={{e}^{i\alpha ({\tau }'',\tau )}}\langle \varphi ,\sigma |{{\varphi }_{{{\mathbf{k}}_{2}}}}{{\varphi }_{{{{\mathbf{{k}}}}_{2}'}}}{{\sigma }_{{{\mathbf{q}}_{2}}}}{{\mathcal{K}}_{{{\mathbf{k}}_{1}},{{{\mathbf{{k}}}}_{1}'},{{\mathbf{q}}_{1}}}}({\tau }',\tau )+\left[ {{\mathcal{K}}_{{{\mathbf{k}}_{1}},{{{\mathbf{{k}}}}_{1}'},{{\mathbf{q}}_{1}}}}({\tau }',\tau ),{{\varphi }_{{{\mathbf{k}}_{2}}}}{{\varphi }_{{{{\mathbf{{k}}}}_{2}'}}}{{\sigma }_{{{\mathbf{q}}_{2}}}} \right]|{{\psi }_{\text{G}}}(\tau )\rangle   \nonumber\\
 & ={{e}^{i\alpha ({\tau }'',\tau )}}{{e}^{i\alpha ({\tau }',\tau )}}{{\varphi }_{{{\mathbf{k}}_{1}}}}{{\varphi }_{{{{\mathbf{{k}}}}_{1}'}}}{{\sigma }_{{{\mathbf{q}}_{1}}}}{{\varphi }_{{{\mathbf{k}}_{2}}}}{{\varphi }_{{{{\mathbf{{k}}}}_{2}'}}}{{\sigma }_{{{\mathbf{q}}_{2}}}}\langle \varphi ,\sigma |{{\psi }_{\text{G}}}(\tau )\rangle   \nonumber\\
 & +{{e}^{i\alpha ({\tau }'',\tau )}}\langle \varphi ,\sigma |\left[ {{\mathcal{K}}_{{{\mathbf{k}}_{1}},{{{\mathbf{{k}}}}_{1}'},{{\mathbf{q}}_{1}}}}({\tau }',\tau ),{{\varphi }_{{{\mathbf{k}}_{2}}}}{{\varphi }_{{{{\mathbf{{k}}}}_{2}'}}}{{\sigma }_{{{\mathbf{q}}_{2}}}} \right]|{{\psi }_{\text{G}}}(\tau )\rangle~,
 \end{align}
where at the third line we have used Equation (\ref{eq:solve K bracket}). The first term appears in the non-Gaussian exponential in Equation (\ref{eq:total wave functional}), and the second term is the correction. For the late time limit $\tau'\to\tau$, we have,
\begin{align}
\mathcal{K}_{\mathbf{k}_1,\mathbf{k}_1',\mathbf{q}_1}(\tau',\tau)\approx\varphi_{\mathbf{k}_1}\varphi_{\mathbf{k}_1'}\sigma_{\mathbf{q}_1}~,
\end{align}
implying the nearly vanishing commutator, and thus the correction is negligible. However, if the intermediate $\tau'$ is distant from the end time $\tau$, the correction cannot be omitted. Thus, the second order correction is mainly contributed to early interaction, and the conformal time cutoff
\begin{align}
-q\tau_{0}=2~,
\end{align}
suppresses this effect. For getting more precise results or studying higher energy theory, the high order terms have to be considered.
\\
\\
Secondly, we have put a Hubble scale UV cutoff for the loop integration. This is a natural choice considering that the sub-horizon modes of the inflaton are in the vacuum and should not decohere the massive field. Nevertheless, it is interesting to study the one-loop regularization and renormalization (counterterms) more carefully.
\\
\\
Finally, considering our formalism is general for massless and massive fields, one can explore more topics related to cosmic quantum decoherence in a wider range. In some results of quantum decoherence in the flat space \cite{Herzog:2002pc,Ho:2013rra}, evidences show that there exists a holographic explanation of quantum decoherence according to AdS/CFT correspondence. Is it applicable to dS/CFT? Can we find a condensed matter analog of quantum decoherence in de Sitter space? We will leave these topics to future research \cite{last}.

\acknowledgments
We thank Hongliang Jiang for his initial collaboration on this topic. We thank Rong-Xin Miao, Elliot Nelson, Bo Ning, Dong-Gang Wang, Zhong-Zhi Xianyu and Yehao Zhou for useful discussions. JL is grateful for PhD assistantships from the departments of physics of Caltech, and warm hosting from department of physics of Hong Kong University of Science and Technology during writing his bachelor's thesis. YW is supported by Grant HKUST4/CRF/13G and ECS 26300316 issued by the Research Grants Council (RGC) of Hong Kong.

\appendix

\section{The effect of boundedness of Hankel functions}\label{app:A}

From Equation (\ref{eq:ready to compute numerically}), the decoherence rate depends on $\mathrm{Re}(\mathcal{Q})$ and a denominator decided by Hankel functions. The Wronskian of mode functions implies that the denominator is time-independent \cite{Burgess:2014eoa}.
\begin{align}
\frac{\partial}{\partial x}\Big[x^{3}\Big|H_{\nu}^{(2)}(x)\Big|^{2}\text{Re}\left(\frac{i}{{{x}^{2}}}\frac{{w}'(x)}{w(x)}\right)\Big]=0.
\end{align}
The denominator may depend on $\nu$, but we only focus on the time dependence which is solely attributed to $\mathrm{Re}(\mathcal{Q})$. For the region of $\frac{\nu^2}{|\nu|}$, the upper bounded of $\mathrm{Re}(\mathcal{Q})$ is estimated.
\begin{align}
&\mathrm{Re}(\mathcal{Q})\leq|\mathcal{Q}|\nonumber \\
&\leq\left(\int_{x}^{2}{d}{x}'\int_{x'}^{2}{d}{x}''\right){{\left({x}'{x}''\right)}^{-\frac{1}{2}}}|H_{\nu}^{(2)}({x}')H_{{{\nu}^{*}}}^{(1)}({x}'')|\int_{1}^{\frac{2}{x''}}{d}{{\mu}_{1}}\int_{-1}^{1}{d{{\mu}_{2}}}\frac{{\left(1-\frac{\mu_{1}^{2}+\mu_{2}^{2}}{2}\right)^{2}}}{{{(\mu_{1}^{2}-\mu_{2}^{2})}^{2}}}\times\nonumber\\
&\sqrt{\left(1+\frac{({{\mu}_{1}}+{{\mu}_{2}})^{2}{x}'^{2}}{4}\right)\left(1+\frac{({{\mu}_{1}}-{{\mu}_{2}})^{2}{x}'^{2}}{4}\right)\left(1+\frac{({{\mu}_{1}}+{{\mu}_{2}})^{2}{x}''^{2}}{4}\right)\left(1+\frac{({{\mu}_{1}}-{{\mu}_{2}})^{2}{x}''^{2}}{4}\right)}\nonumber\\
&\leq25\left(\int_{x}^{2}{d}{x}'\int_{x'}^{2}{d}{x}''\right){{\left({x}'{x}''\right)}^{-\frac{1}{2}}}|H_{\nu}^{(2)}({x}')H_{{{\nu}^{*}}}^{(1)}({x}'')|\int_{1}^{\frac{2}{x''}}{d}{{\mu}_{1}}\int_{-1}^{1}{d{{\mu}_{2}}}\frac{{\left(1-\frac{\mu_{1}^{2}+\mu_{2}^{2}}{2}\right)^{2}}}{{{(\mu_{1}^{2}-\mu_{2}^{2})}^{2}}}\nonumber\\
&=25\left(\int_{x}^{2}{d}{x}'\int_{x'}^{2}{d}{x}''\right){{\left({x}'{x}''\right)}^{-\frac{1}{2}}}|H_{\nu}^{(2)}({x}')H_{{{\nu}^{*}}}^{(1)}({x}'')|\times\nonumber\\
&\Bigg\{\frac{-\left(x''^{4}-4x''^{2}+16\right)\tanh^{-1}\left(\frac{x''}{2}\right)+2x''\Big[x''(x''+i\pi-6)+2x''\tanh^{-1}\left(\frac{2}{x''}\right)+8\Big]}{8x''{}^{2}}\Bigg\}\nonumber\\
&=25\left(\int_{x}^{2}{d}{x}'\int_{x'}^{2}{d}{x}''\right){{\left({x}'{x}''\right)}^{-\frac{1}{2}}}|H_{\nu}^{(2)}({x}')H_{{{\nu}^{*}}}^{(1)}({x}'')|\Big(\frac{1}{x''}+\mathcal{O}(1)\Big)\nonumber\\
&\leq B'\int_{x}^{2}{d}{x}'\int_{x'}^{2}{d}{x}''x'^{-\frac{1}{2}}x''^{-\frac{3}{2}}\nonumber\\
&=2B'\Big(\mathrm{log}(\frac{1}{x})+\mathcal{O}(1)\Big)~.
\end{align}
In the third inequality, we apply the conditions $x'\leq x''$, $\mu_{2}\leq\mu_{1}$
and $\mu_{1}x''\leq2$. Also, the integrals of $\mu_{1}$ and $\mu_{2}$
are expanded in order to estimate the late-time divergence. For imaginary
$\nu$, the absolute values of Hankel functions are bounded, namely
$|H_{\nu}^{(2)}({x}')H_{{{\nu}^{*}}}^{(1)}({x}'')|\leq\frac{B'}{25}$,
which is shown in the last inequality. Therefore, the divergence of
$M(\nu,x)$ is no faster than logarithm at late time, and this can
explain why the decoherence rate increases slowly in the region with
$\frac{\nu^{2}}{|\nu|}<0$. On the other hand, the Hankel functions cause polynomial divergence at late time in the region with $\frac{\nu^{2}}{|\nu|}>0$, leading the rapid change of decoherence rate.
\\
\\
Numerically, the upper bound of the delay for decoherence with imaginary $\nu$ can
be estimated by linear approximation
\begin{align}
M(\nu,x) \lesssim k_{\nu}\mathrm{log}(\frac{1}{x})+b_{\nu}=k_{\nu}N(x)+b_{\nu}~,
\end{align}
where $k_{\nu}$ is the increment per e-fold, and 
\begin{align}
N\gtrsim \frac{M(\nu,x)-b_{\nu}}{k_{\nu}}\approx\frac{\mathcal{O}(10^{7}\sim10^{13})}{k_{\nu}}~.
\end{align}

\end{document}